\documentclass[12pt]{article}
  \usepackage{epsfig}
  \usepackage{psfrag}
  \usepackage{a4}

  \newcommand{\be}{\begin{equation}}
  \newcommand{\ee}{\end{equation}}

\begin{document}

  \title{Reply to Johansen's comment}

  \author{L. Laloux, M. Potters,\\ 
  J.P. Aguilar, J.P. Bouchaud\\
 }

\maketitle

Our paper [1] contained a series of comments on the claims that were then made
about possible log-period precursors to financial crashes. We felt in particular 
that a seven parameter fit of noisy data, without any theoretical guidance,
was dangerous; that the core of such a spectacular effect, i.e. the 
geometrical acceleration of the periods of the oscillations should, if present, 
be visible to the naked eye. We expressed doubts about the very nature
of the prediction: log periodic singularities are interesting if there
{\it is} a singularity such as a {\it sudden} drop of price, but in many examples, 
the observed crash occurs before
the predicted singularity (as in November 1997 [1]), or actually does not
occur as a sudden crash but as a smooth drawdown, as now argued by Johansen [2] for the 
{\sc jgb} (Japanese Governement Bonds) in 1995. The mechanism that would lead to such a 
variety of scenarii, 
and still preserve to a `log periodic' signal with no singularity, 
would be, to say the least, non standard. 
			
The situation has evolved in the recent years. More statistical tests and
evidence were presented by Johansen and Sornette (JS) [3]. Interestingly, the parameter 
governing the acceleration of the oscillations was found to be rather constant across 
many different crashes. However, strong and
well documented criticisms were also expressed by Feigenbaum in two papers [4] that are,
sadly, rarely cited by JS.

Finally, in [1], we reported a rather anectodal event, that we felt relevant because 
it was at the time a real {\it prediction}, rather than a `post-diction'. As we 
argued then, the failure of a prediction does not prove the theory to be wrong. Our 
point, however, was that {\it both} failures and success should have been 
systematically reported. The two specific points of the authors comment concern 
(i) the predicted
date of the crash of the {\sc jgb} in 1995 and (ii) a discussion about whether or
not the crash did actually happen. About point (i), the author is perfectly
right. We have found the precise date
of the trades reported in [1], that start on the 13th of July 1995. As we wrote in [1], 
this prediction was first reported to us in May, but the prediction was indeed for 
August, and obviously not for May itself, as we mistakenly wrote.  
About point (ii), the argument of Johansen is that although the {\sc jgb} did not ``crash'' in
August 1995, its subsequent drawdown can be seen as a crash, thereby validating 
a posteriori the log-periodic scenario, since its amplitude is ``large'' --
in particular for a bond contract, ``less volatile'' than stock markets. That the
decline is `large' is infered from the from the fact that its amplitude lies above the
extrapolation of an ad-hoc stretched exponential fit of the distribution of 
``normal'' events. However, there is no convincing theoretical model underpinning this 
particular functional form. On the contrary, both precise 
empirical studies [5] and recent rather successful theoretical models [6] 
suggest a much fatter, 
power-law tail, for which the observed drawdown would not stand as an outlier. We 
furthermore note that
bond markets {\it do} crash in a usual sense: for example, on October 8 and 9, 1998 the 
(December) Bund futures lost respectively 1.96 and 1.53 point, a 6.7 and a 5.2 sigma
event (as mesured by the historical volatility since 1997). These numbers are comparable 
to the worst crashes of the last ten years on the S\&P 500, Aug 31 1998 and Nov 27 1997, 
two 6-sigma events. It can also be seen on Johansen's Fig. 1 that the amplitude of the drawdown 
is actually less than the amplitude of the `draw-up' that occured just before. Should there
then be log-periodic oscillations before this draw-up? 

Finally, we would like to remind Johansen that `put' options are worth nothing if the 
contract is above
the exercise price at maturity. Delicate trading {\it was} necessary, 
not because the {\sc jgb} dropped,
but precisely because a crash in the usual sense did not occur.

\vskip 1cm

  {\bf The authors are at Science \&
  Finance, the research division of Capital Fund
  Management, 109-111 rue Victor Hugo, 92523 Levallois {\sc cedex}, France. 
 Jean-Philippe Bouchaud is also at the Service de
  Physique de l'Etat Condens\'e, CEA Saclay, Orme des Merisiers, 91191 Gif sur Yvette {\sc cedex}, 
France.}

\vskip 1cm

[1] L. Laloux, M. Potters, R. Cont, J.P. Aguilar, J.P. Bouchaud, Europhys. Lett. {\bf 45},
1 (1999)        

[2] A. Johansen, preceeding comment.

[3] D. Sornette, A. Johansen, Quantitative Finance, {\bf 1} 452 (2001)

[4] J.A. Feigenbaum, Quantitative Finance, {\bf 1} 346 (2001), ibid. {\bf 1} 527 (2001).

[5] V. Plerou, P. Gopikrishnan,  L.A. 
Amaral, M. Meyer, H.E. Stanley, Phys. Rev. {\bf E60} 6519 (1999).

[6] J.-F. Muzy, J. Delour, E. Bacry, Eur. Phys. J. B 17, 537-548 (2000)

  \end{document}